\documentclass[twocolumn,showpacs,showkeys,pra,smaller,amssymb,amsfonts,amsmath,
superscriptaddress]{revtex4}
\usepackage{graphicx}
\usepackage[utf8]{inputenc}
\usepackage[T1]{fontenc}
\usepackage[colorlinks,linkcolor=black,citecolor=blue]{hyperref}
\usepackage{float}
\newcommand{\opn}[1]{\ensuremath{\operatorname{#1}}} 
\begin{document}

\preprint{Physical Review A}
\title{Robustness of continuous-variable entanglement via geometrical 
nonlinearity}
\author{Philippe Djorwé}
\email{djorwepp@gmail.com}
\affiliation{Laboratory of Modelling and Simulation in Engineering, Biomimetics
and Prototypes, Department of Physics, Faculty of Science,\\  University of 
Yaoundé I, P.O. Box 812, Yaoundé, Cameroon}
\author{S.G. Nana Engo}
\email{snana@univ-ndere.cm}
\affiliation{Laboratory of Photonics, Department of Physics, Faculty of Science,
University of Ngaoundéré, P.O. Box 454, Ngaoundéré, Cameroon}
\author{Paul Woafo}
\email{pwoafo1@yahoo.fr}
\affiliation{Laboratory of Modelling and Simulation in Engineering, Biomimetics
and Prototypes, Department of Physics, Faculty of Science,\\  University of 
Yaoundé I, P.O. Box 812, Yaoundé, Cameroon}
\date{\today}

\begin{abstract}
We propose a scheme to generate robust stationary continuous-variable 
entanglement in optomechanical systems, based on geometrical nonlinearity that 
occurs for large mechanical displacements. Such nonlinearity is often used to 
correct the dynamics of the systems in the strong coupling regime. It appears 
that geometrical nonlinearity enhances the entanglement and shifts its maximum 
towards high detuning values. Using the experimental parameters, we find that 
such a scheme generates a very robust entanglement against thermal decoherence 
even at room temperature. Our results show that geometrical nonlinearity affects 
entanglement as the optomechanical quantum interface. 
\end{abstract}

\maketitle

\section{Introduction}
After success on the quantum ground state achievable \cite{1,2}, the current 
challenge now is to know how far quantum theory, bizarre as it is, can allow the 
observation of quantum effects \cite{3}. A suitable effect that can inform us 
about this quantum theory's limit is quantum entanglement, since it is a 
distinguishing feature that separates quantum from classical physics. Quantum 
entanglement plays an important role in sharing information in quantum networks 
\cite{4,5,6}. In Particular, the continuous-variable (CV) entangled quantum 
states constitute a key element to challenge quantum communication processes 
such as quantum cryptography \cite{7}, quantum teleportation \cite{8,9,10}, 
quantum dense coding \cite{11}, and quantum computational tasks \cite{12}. 

Many theoretical and experimental schemes have been proposed to generate CV 
entangled quantum states \cite{13,14,15,16,17,18}. According to the great 
interest in the nanomechanics during this past decade, optomechanical systems 
\cite{9,11,19,20,21,22}, superconducting microwave cavities \cite{23,24}, and 
hybrid systems \cite{25,26} have been very recently used as a tool for 
generating strong CV entangled quantum states. This consists in coupling the  
optical and/or microwave mode to the mechanical mode of the resonator that 
vibrates under the electromagnetic field. There are many configurations that 
have been proposed to this end. In \cite{19}, stationary entanglement between 
the optical field and the mechanical mode of a vibrating end mirror is generated 
from the optical Fabry-Pérot cavity. Strong quantum correlation between the 
mirror and the optical Stokes sideband is explained in \cite{20} as being 
generated by a scattering process. The configuration used in \cite{21} consists 
of a whispering-gallery mode cavity with a movable boundary. Entanglement 
between the mechanical and radiation modes is achieved in \cite{22} by using a 
suitable modulation of the driving field. A pulsed field is used to create 
Einstein-Podolsky-Rosen-type entanglement between the mechanical mode and light 
pulses and  microwave pulses respectively in optomechanical cavity \cite{9} and 
a microwave cavity, respectively \cite{24}. In \cite{11}, a 
membrane-in-the-middle geometry is used to generate output entangled light from 
a fixed end Fabry-Pérot cavity. The optical-microwave quantum interface is used 
in \cite{25,26} to a produce robust entangled signal that can be used for 
high-fidelity transfer of quantum states between optical and microwave fields. 
All these schemes aim to generate robust entangled states against decoherence, 
which limits their lifetime \cite{5} and their performance in quantum 
applications. This decoherence is often manifested by various factors such as 
the stability conditions that place constraints on the magnitude of the 
effective optomechanical couplings and the thermal noise of the mechanical 
mode. 

Recent experimental studies have shown that a cantilever nanobeam  can exhibit 
both stiffening ($+$) and softening ($-$) geometrical nonlinearity 
\cite{27,28}. This intrinsic anharmonicity in the mechanical motion of 
micromechanical and nanomechanical resonators is usually small and therefore 
only relevant in the regime of large oscillation amplitudes where its 
contribution becomes important \cite{27,28}. In order to detect quantum 
behavior, it is a common approach to introduce nonlinearities in the quantum 
system \cite{29}. It is noteworthy that the dynamics of the purely harmonic 
quantum system is analogous to its classical dynamics, in the sense that 
expectation values of canonical observables follow the classical equations of 
motion. Indeed, stiffening geometrical nonlinearity has being used to squeeze 
the mechanical mode in optomechanical systems \cite{30} and for both quantum 
control and quantum information processing \cite{29} whereas softening 
nonlinearity is shown as a factor that limits some quantum effects in 
optomechanics \cite{31,32}.

In this paper, motivated by the recent experimental parameters of Ref. 
\cite{1}, we propose a scheme that generates robust CV entanglement against 
thermal decoherence. This consists to take into account the geometrical 
nonlinearity which makes a correction on the dynamics of the resonator, in the 
limit of large displacements \cite{27,28,29,30,31}. Such a scheme can be 
extended to superconducting microwave circuits and hybrid systems \cite{26}. 
Section \ref{sec:Model} describes the system used and details the mathematical 
tools for the entanglement. We then show, in Sec. \ref{sec:Eff} how the 
geometrical nonlinearity is useful to enhance the entanglement even at room 
temperature. We summarize in Sec. \ref{sec:Concl}.

\section{Model and dynamics equations}\label{sec:Model}

We study a Fabry-Pérot optomechanical cavity, in which one of the cavity’s
mirrors is free to move like a flexural mirror cantilever beam. This system is
described, in the rotating-wave approximation picture with respect to
$\hbar \omega_{\ell}\alpha ^{\dag }\alpha $, by the single-mode Hamiltonian 
\cite{33}, 
\begin{small}
\begin{equation}
\begin{split}
H&=\hbar(\Delta _0-g_mx_m) \alpha ^{\dag }\alpha +\frac{\hbar\Omega _m}{4}
\left( p_m^2+x_m^2-\frac{\beta'x_{ZPF}^2}{2\Omega _m^2}x_m^4\right) \\
&+\hbar E_0\left( \alpha+\alpha ^{\dag }\right).  \label{eq:a}
\end{split}
\end{equation}
\end{small}
Here $x_m$ and $p_m$ are the dimensionless position and momentum operators of 
the mechanical oscillator related to their counterparts operators of the 
nanobeam as $x=x_{ZPF}x_m$ and $p=\frac{\hbar}{x_{ZPF}}p_m$, with $[x_m,p_m]=2i$ 
and $x_{ZPF}=\sqrt{\frac{\hbar}{2M\Omega _m}}$. $\Omega_m$ and $M$ are the 
mechanical frequency and the effective mass of the mechanical mode. The 
annihilation $\alpha$ and creation $\alpha^{\dag }$ photon operator are defined 
with respect to $[\alpha,\alpha^{\dag}]=1$. The first term of the Hamiltonian 
describes both the energy of the cavity mode and the radiation pressure coupling 
of the rate $g_m=\sqrt{2}\omega_c\frac{x_{ZPF}}{d_0}$, where $\omega_c$ and 
$d_0$ are the cavity frequency and cavity length, respectively. The second term 
is the anharmonic mechanical energy of the nanoresonator with 
$\frac{\beta'x_{ZPF}^2 x_m^4}{2\Omega_m^2}$ the anharmonic term. The third term 
is the input driving by a laser with frequency $\omega_{\ell}$, where the 
amplitude of the input laser beam $E_0$ is related to the input laser power 
$P_0$ by $|E_0|= \sqrt{\frac{2P_0\kappa}{\hbar \omega_{\ell}}}$. The 
laser-cavity detuning is defined as $\Delta_0=\omega_{\ell}-\omega_c$. The 
aforementioned nonlinear term occurs in the system when the nanoresonator is 
subjected to large displacement amplitudes. This comes about because the 
flexure causes the beam to lengthen, which at large amplitudes adds a 
significant correction to the overall elastic response of the beam 
\cite{27,28,29,30,31}. Depending on the sign of this nonlinearity, one 
distinguishes the softening ($-$) and the stiffening ($+$) nonlinearity which 
both merge in the nanostructure dynamics. Conceptually, a doubly clamped beam 
exhibits a stiffening nonlinearity and the cantilever has both stiffening and 
softening nonlinearity \cite{27,28}. The sign of the nonlinearity depends also 
on the material nonlinearity. Materials become either stiffer of softer for 
large strains. At some large deflection, the nanodevices will become affected by 
the material nonlinearity (further informations are available in \cite{27,28}). 
Here, we take into account a softening nonlinearity which merges in the 
optomechanical coupling between the radiation pressure and the position of the 
flexural cantilever mirror. The parameters used here, given in Table 
\ref{tab:ExpData}, are those of the system that was recently studied 
experimentally in Ref. \cite{1}.
\begin{table}[htbp]
 \centering
\begin{tabular}{c|c|c|c|c|c|c}\hline\hline
 $\Omega_m/2\pi$ & $\Gamma_{m}/2\pi$ & $Q_{m}$ & $g_0/2\pi$ &
$\kappa/\pi$& $P_0$& $T$\\\hline$3.6\opn{GHz}$ & $35\opn{kHz}$ &
$1.05\times10^{5}$ &$910\opn{kHz}$ & $529\opn{MHz}$ & $0.7\opn{mW}$&
$270\opn{mK}$\\\hline\hline
\end{tabular}
 \caption{Experimental parameters of Ref. \cite{1} used in this work.}
 \label{tab:ExpData}
\end{table}

Considering the photon losses in the cavity and the Brownian noise acting on
the resonator, one derives from the Hamiltonian (\ref{eq:a}), the nonlinear 
quantum Langevin equations (QLEs) \cite{31} 
\begin{subequations}
\label{eq:1}
\begin{align}
\dot{x}_m& =\Omega _mp_m, \\
\dot{p}_m& =-\Omega _mx_m-\Gamma _mp_m+g_m\alpha ^{\dag }\alpha+\beta''+F_{th}, 
\\
\dot{\alpha}& =\left[ i(\Delta _0+g_mx_m)-\frac{\kappa }{2}\right]
\alpha -iE_0+\sqrt{\kappa}\alpha ^{in},
\end{align}
\end{subequations}
where $\beta''=\frac{\beta'x_{ZPF}^2}{\Omega_m}x_m^3$. We have introduced 
both the vacuum radiation input noise operator $\alpha^{in}$, whose only 
nonzero correlation function is
\begin{equation}
\langle\alpha^{in}(t)\alpha^{in,\dag}(t')\rangle =\delta (t-t'),
\end{equation}
and the Hermitian Brownian noise operator $F_{th}$, with the correlation 
function
\begin{equation}
\langle F_{th}(t)F_{th}(t')\rangle =\frac{\Gamma _m}{\Omega _m}\int 
\frac{d\Omega}{2\pi}e^{-i\Omega(t-t')}\Omega \left[ \coth \left( \frac{\hbar 
\Omega}{2k_BT}\right) +1\right],  \label{eq:aa}
\end{equation}
where $k_B$ the Boltzmann constant and $T$ the bath temperature.

By setting the time derivatives to zero in the set of nonlinear 
equations (\ref{eq:1}), the steady-state values yield $p_s=0,x_s=2\frac{g_m} 
{\Omega_m}n_s$ and $|E_0|^2=n_s(\Delta^2+\frac{\kappa ^2}{4})$, 
where we have set the steady state photon number in the cavity equal to 
$n_s=|\alpha_s|^2$. Here $\Delta=\Delta_0+2\frac{g_m^2}{\Omega_m}n_s$ is the
effective detuning. One can rewrite each Heisenberg operator as a $c$-number
steady-state value plus an additional fluctuation operator with zero-mean value,
$p_m=p_s+\delta p_m$, $x_m=x_s+\delta x_m$, and $\alpha =\alpha_s+\delta \alpha 
$. Inserting these expressions into the set of QLEs equations (\ref{eq:1}), 
leads to a set of quantum Langevin equations for the fluctuation operators. As 
the parameter regime relevant for generating optomechanical entanglement is 
that with a very large input power $P_0$, i.e., for strong coupling 
$|\alpha_s|\gg 1$, one can safely neglect the higher order terms of fluctuations
\cite{19}.

By introducing the vector of quadrature fluctuations $u(t)=(\delta 
x_m(t),\delta p_m(t),\delta I(t),\delta \varphi (t))^T$ and the vector of 
noises $n(t)=(0, F_{th}(t),\sqrt{\kappa}\delta I^{in}(t),\sqrt{\kappa}\delta 
\varphi^{in}(t))^T$ \cite{32}, the linearized QLEs can be written in compact 
form 
\begin{equation}
\dot{u}(t)=Au(t)+n(t),  \label{eq:b}
\end{equation}
where 
\begin{equation}
A=\begin{pmatrix}
0 & \Omega _m & 0 & 0 \\ 
\Omega _m(\beta -1) & -\Gamma _m & G & 0 \\ 
0 & 0 & -\frac{\kappa}{2} & -\Delta \\ 
G & 0 & \Delta & -\frac{\kappa}{2}
\end{pmatrix}.
\end{equation}
The term $G=g_m|\bar{\alpha}|$ is the effective optomechanical coupling that 
can generate a significant entanglement in the strong-coupling limit. The 
parameter $\beta =\frac{3\beta' x_s^2}{\Omega_m^2}$ is the dimensionless 
geometrical nonlinearity that is used in the following (refer to \cite{30,31}
for more detailed calculations). 

By integrating Eq. (\ref{eq:b}), we obtain
\begin{equation}
u(t)=M(t)u(0) +\int_0^tdsM(s)n(s),\label{eq:c}
\end{equation}
where $M(t)=\exp(At)$. To analyze the stability of our system, we have used the 
Routh-Hurwitz criterion which yield the two nontrivial conditions on the system 
parameters
\begin{align}
\begin{split}
s_1&=\Gamma_m\kappa\left\{\left[\frac{\kappa^2}{4}+(\Omega_m-\Delta)^2\right] 
\left[ \frac{\kappa ^2}{4}(\Omega_m+\Delta)^2\right]\right. \\
& \left. +\Gamma_m\left[(\Gamma_m+\kappa)\left(\frac{\kappa^2}{4}+\Delta 
^2\right)+\kappa\Omega_m^2\right] \right\}\\
& -\Delta\Omega_mG^2(\Gamma_m+\kappa)^2>0,\label{eq:d}
\end{split}\\
s_2&=\Omega_m\left(\Delta^2+\frac{\kappa ^2}{4}\right) +G^2\Delta >0.
\label{eq:e}
\end{align}
With respect to physical parameters in Table \ref{tab:ExpData}, these stability 
conditions lead at the blue sideband of the cavity ($\Delta =-\Omega_m$) to 
$G<2.26\times 10^{10}s^{-1}$ or $\alpha_s<2.07\times 10^{3}$ while at the red 
sideband ($\Delta =\Omega_m$) one has $G<2.89\times10^{7}s^{-1}$ or 
$\alpha_s<0.26$.  Optomechanical entanglement is then possible within the 
blue-sideband detuning range where the coupling is strong. Due to Eq. 
(\ref{eq:aa}), the mirror Brownian noise $F_{th}(t)$ is not $\delta$ correlated 
and therefore does not describe a Markovian process. However, as the parameters 
used here lead to a large mechanical factor $Q=\frac{\Omega_m}{\Gamma_m}\approx 
1.05\times 10^5\gg 1$, the quantum effects can be achieved. This limit allows 
$F_{th}(t)$ to be $\delta$ correlated \cite{19}, 
\begin{equation}
\frac{1}{2}\langle F_{th}(t)F_{th}(t')+F_{th}(t')F_{th}(t)\rangle 
\approx \Gamma _m(2n_{th}+1)\delta (t-t'),
\end{equation}
where the mean thermal excitation number is set equal to $n_{th}=(e^{\hbar 
\Omega_m/ k_BT}-1)^{-1}$ and the Markovian process can then be recovered. 

Since the quantum noises $F_{th}$ and $\alpha ^{in}$ are zero-mean quantum 
Gaussian noises and the dynamics is linearized, the quantum steady state for 
the fluctuations is a zero-mean bipartite Gaussian state, fully characterized 
by its $4\times 4$ correlation matrix which has the components 
$V_{ij}=\frac{1}{2} \langle u_i(\infty)u_j(\infty)+u_j(\infty) u_i(\infty) 
\rangle$. When the system is stable, using Eq.(\ref{eq:c}), we
obtain
\begin{equation}
V_{ij}=\sum_{kl}\int_0^{\infty }ds\int_0^{\infty }ds'M_{ik}(s)M_{jl}(s') \Phi 
_{kl}(s-s'), \label{eq:f}
\end{equation}
where $\Phi_{kl}(s-s')=\frac{1}{2}\langle n_k(s)n_{\ell}(s')+n_{\ell}(s')n_k(s) 
\rangle$ is the matrix of stationary noise correlation functions. Using the 
fact that the three components of $n(t)$ are uncorrelated, we find 
$\Phi_{kl}(s-s') =D_{kl}\delta(s-s')$, where 
$D=\mathrm{diag}[0,\Gamma_m(2n_{th}+1), 
\kappa,\kappa] $ is a diagonal matrix and Eq. (\ref{eq:f}) becomes 
$V=\int_0^{\infty} dsM(s)DM(s)^T$,  which leads, for a stable system 
($M(\infty) =0$) and after applying Lyapunov's first theorem \cite{34}, to
\begin{equation}
AV+VA^T=-D.  \label{eq:g}
\end{equation}
Here $V$ is known as the system covariance matrix (CM) and it contains all 
information about the steady state. Equation (\ref{eq:g}) can be 
straightforwardly solved, but the general exact expression is too cumbersome 
and will not be reported here. 

\section{Effects of geometrical nonlinearity on stationary entanglement}
\label{sec:Eff}

The entanglement of the steady state can then be quantified by means of the 
logarithmic negativity $E_{\mathcal{N}}$ \cite{15} which, in the CV case, is 
defined as \cite{15,16}
\begin{equation}
E_{\mathcal{N}}=\max[0,-\ln 2\eta],  \label{eq:h}
\end{equation}%
where 
\begin{equation}
\eta =\sqrt{\frac{\sum (V) -\sqrt{\sum (V)^2-4\det V}}{2}},
\end{equation} 
is the lowest symplectic eigenvalue of the partial transpose of the CM, with 
$\sum(V)$ expressed in terms of the $2\times 2$ block matrix
\begin{equation}
V=\begin{bmatrix}
V_m & V_{corr} \\ 
V_{corr}^T & V_{cav}
\end{bmatrix},
\end{equation}
as $\sum(V)=\det V_m+\det V_{cav}-2\det V_{corr}$. The matrix $V_m$ is 
associated with the oscillating mirror, $V_{cav}$ is associated with the cavity 
mode, and $V_{corr}$ describes the optomechanical correlations. According to 
Eq.(\ref{eq:h}), a Gaussian state is entangled ($E_{\mathcal{N}}>0$) if and only 
if $\eta <\frac{1}{2}$, which is equivalent to Simon's necessary and sufficient 
entanglement nonpositive partial transpose criterion for Gaussian states 
\cite{16}. 

\begin{figure*}[tbp]
\centering
\begin{minipage}{16cm}
\begin{center}
\resizebox{0.45\textwidth}{!}{
\includegraphics{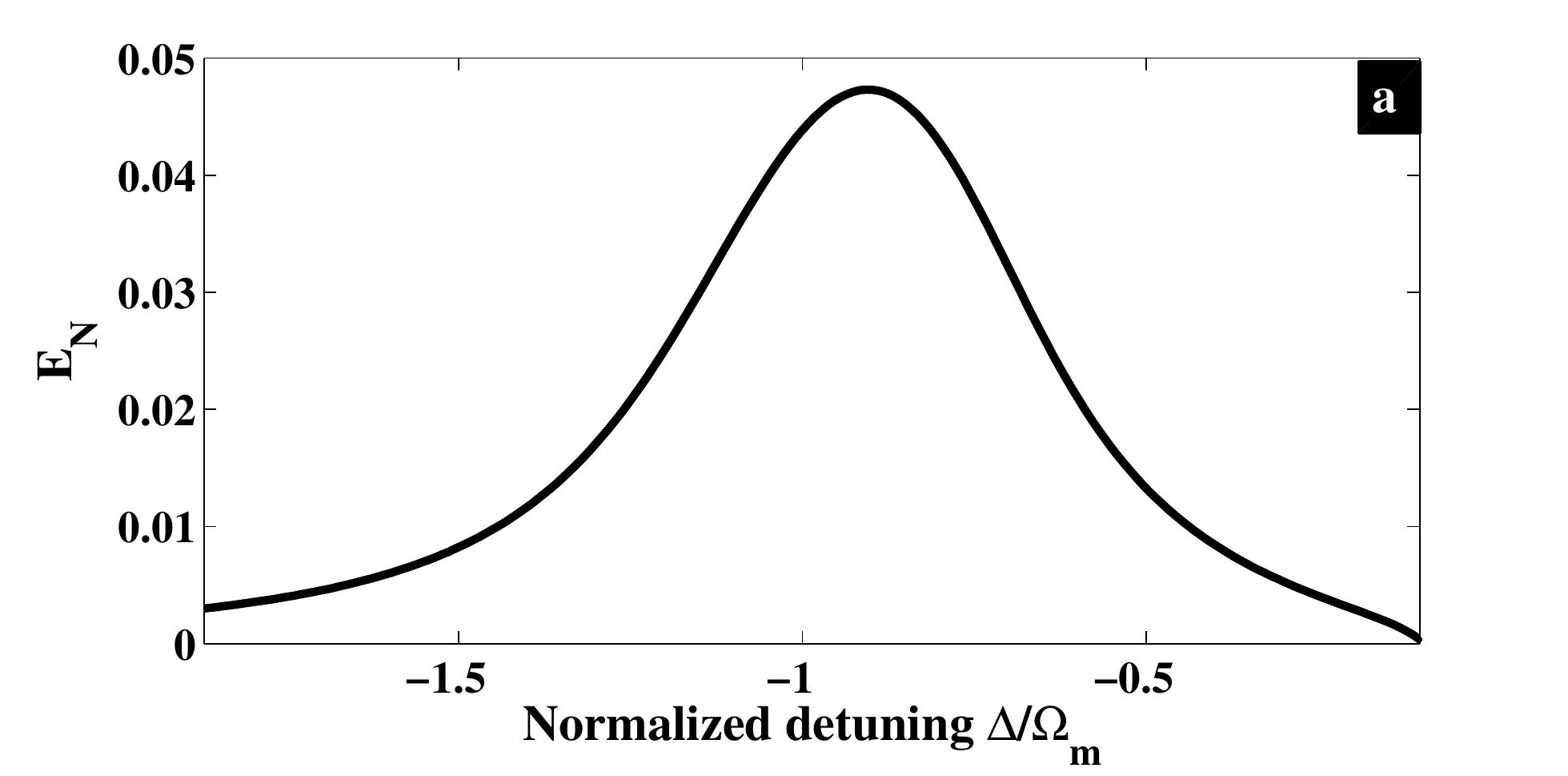}}
\resizebox{0.45\textwidth}{!}{
\includegraphics{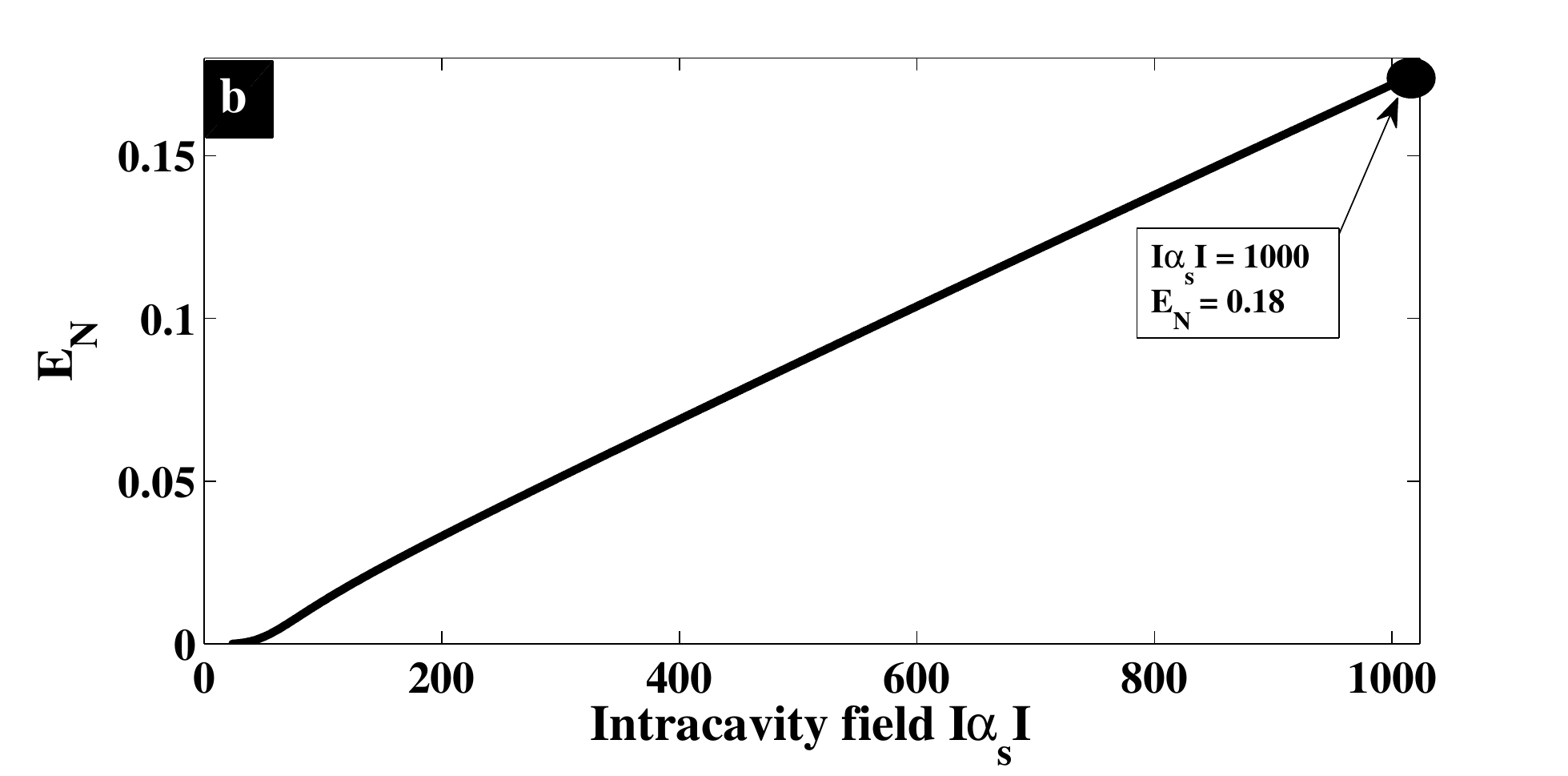}}
\caption{Linear regime obtained with the physical parameters in 
Table \ref{tab:ExpData}. (a) Logarithmic negativity $E_{\mathcal{N}}$ versus 
normalized detuning $\frac{\Delta}{\Omega_m}$. The entanglement is maximal 
around the blue-detuning $\frac{\Delta}{\Omega_m}=-1$. (b) Logarithmic 
negativity $E_{\mathcal{N}}$ versus intracavity field with high input power 
$P_0=10\opn{mW}$. The entanglement is enhanced for high input power.}
\label{fig:Fig1} 
\end{center}
\end{minipage}    
\end{figure*}

The parameters of Table \ref{tab:ExpData} not only satisfy our aforementioned 
stability conditions, but also lead to a generation of a lightly mechanical and 
optical mode entanglement in linear regime. This entanglement, which is present 
only in a finite interval of detuning values around the blue-sideband 
$\frac{\Delta}{\Omega_m}=-1$ [Fig.\ref{fig:Fig1}(a)], can be increased by 
significant optomechanical coupling as shown in \cite{19,21,23,26}, which 
consists in increasing,for example, the input power $P_0$. This is shown in 
Fig. \ref{fig:Fig1}(b), where the high power is $P_0=10\opn{mW}$, which 
corresponds to the intracavity field of $|\alpha_s| =10^{3}$, and is in 
accordance with both current state-of-the-art optics and our stability 
condition. The above value of the intracavity field has been recently used to 
generate entangled states \cite{21} and squeezed states \cite{26} in 
optomechanical systems. Such a high input power induces large mechanical 
displacement of the nanoresonator, which exhibits a geometrical nonlinearity in 
the system \cite{29,30,31}.

\begin{figure*}[tbp]
\centering
\begin{minipage}{16cm}
\begin{center}
\resizebox{0.45\textwidth}{!}{
\includegraphics{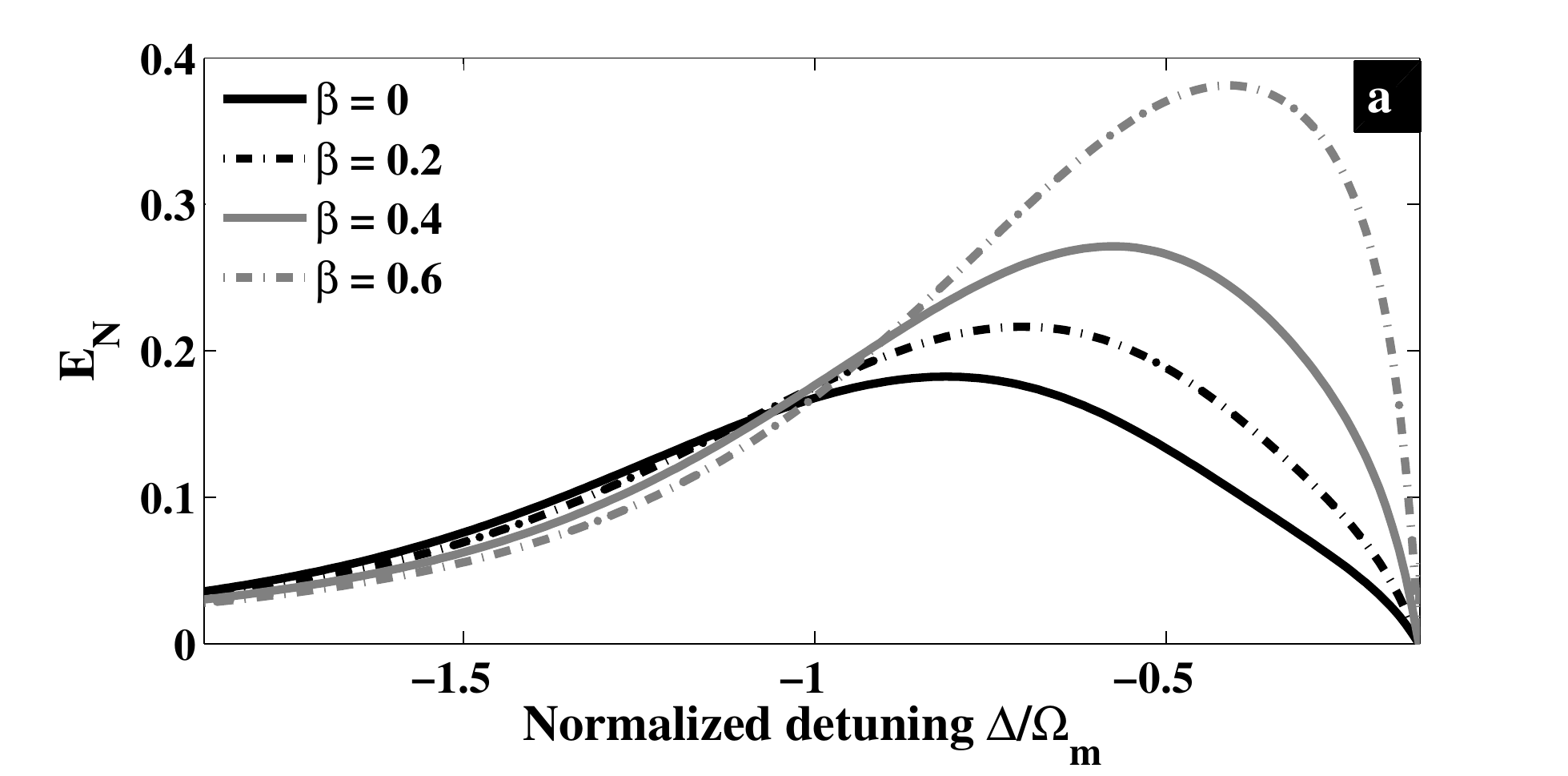}}
\resizebox{0.45\textwidth}{!}{
\includegraphics{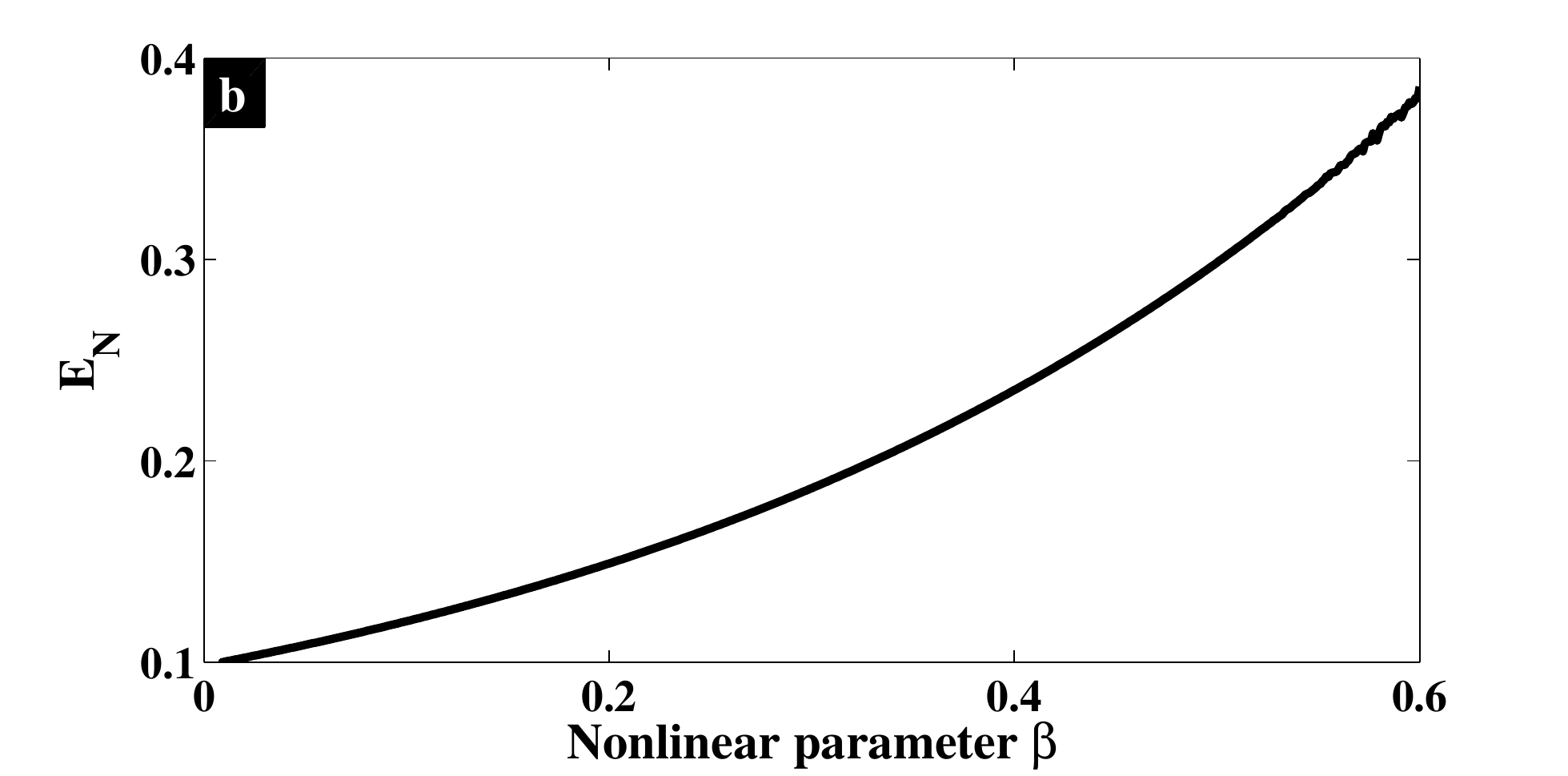}}
\caption{Plots of the input power $P_0=10\opn{mW}$ and the others parameters 
are those in Table \ref{tab:ExpData}. (a) Logarithmic negativity versus the 
normalized detuning for different values of geometrical nonlinearity. The 
increase of nonlinear parameter enhances the entanglement and shifts its 
maximum value towards high detuning values (compare the gray dash-dotted line 
and solid black line). (b) Logarithmic negativity versus the geometrical 
nonlinearity for $\frac{\Delta}{\Omega_m} =-0.5$. This shows the robustness of 
the entanglement in the presence of geometrical nonlinearity.}
\label{fig:Fig2} 
\end{center}
\end{minipage}    
\end{figure*}

A more interesting situation is depicted in Fig. \ref{fig:Fig2} (a), which 
represents the logarithmic negativity versus the normalized detuning for 
different values of the nonlinear parameter $\beta$. We remarks that the 
entanglement increment is related to an increase of the nonlinear parameter 
for $\frac{\Delta}{\Omega_m}\in[-1.1;-0.16]$, with a significant enhancement 
around the normalized detuning $\frac{\Delta}{\Omega_m}=-0.5$. In other terms, 
the entanglement becomes robust with the nonlinearity within the above interval 
of $\frac{\Delta}{\Omega_m}$. In the case where $\beta=0$ (linear case), the 
optimal entanglement remains near $\Delta =-\Omega_m$ (solid black line in 
Fig. \ref{fig:Fig2}(a) and for $\beta\neq 0$ it is shifted towards high 
detuning values [other curves in Fig.\ref{fig:Fig2} (a)]. Reverse effects are 
induced on entanglement in \cite{19} and \cite{35} by the resonator mass and the 
Kerr nonlinearity respectively. This means that the softening nonlinear effect 
study here is a promising way to improve quantum information processing such as 
quantum teleportation and quantum key distribution. For the small normalized 
detuning values, $\frac{\Delta}{\Omega_m}\in[-2;-1.1]$, the increase of the 
nonlinearity induces a decrease of the entanglement. However, this opposite 
effect of the parameter $\beta$ on the entanglement is not important in the 
system. In Fig. \ref{fig:Fig2} (b) we have set the normalized detuning equal 
to $\frac{\Delta}{\Omega_m}=-0.5$ and varied the nonlinear parameter $\beta$. 
This curve shows the robustness of entanglement when the nonlinearity increases 
in the system. Thus, in the system that exhibits geometrical nonlinearity, the 
optimal entanglement depends on the value of the detuning and should be checked 
around the detuning $\frac{\Delta}{\Omega_m}=-0.5$ but not around 
$\frac{\Delta}{\Omega_m}=-1$ as in linear system \cite{19}.

\begin{figure*}[tbp]
\centering
\begin{minipage}{16cm}
\begin{center}
\resizebox{0.45\textwidth}{!}{
\includegraphics{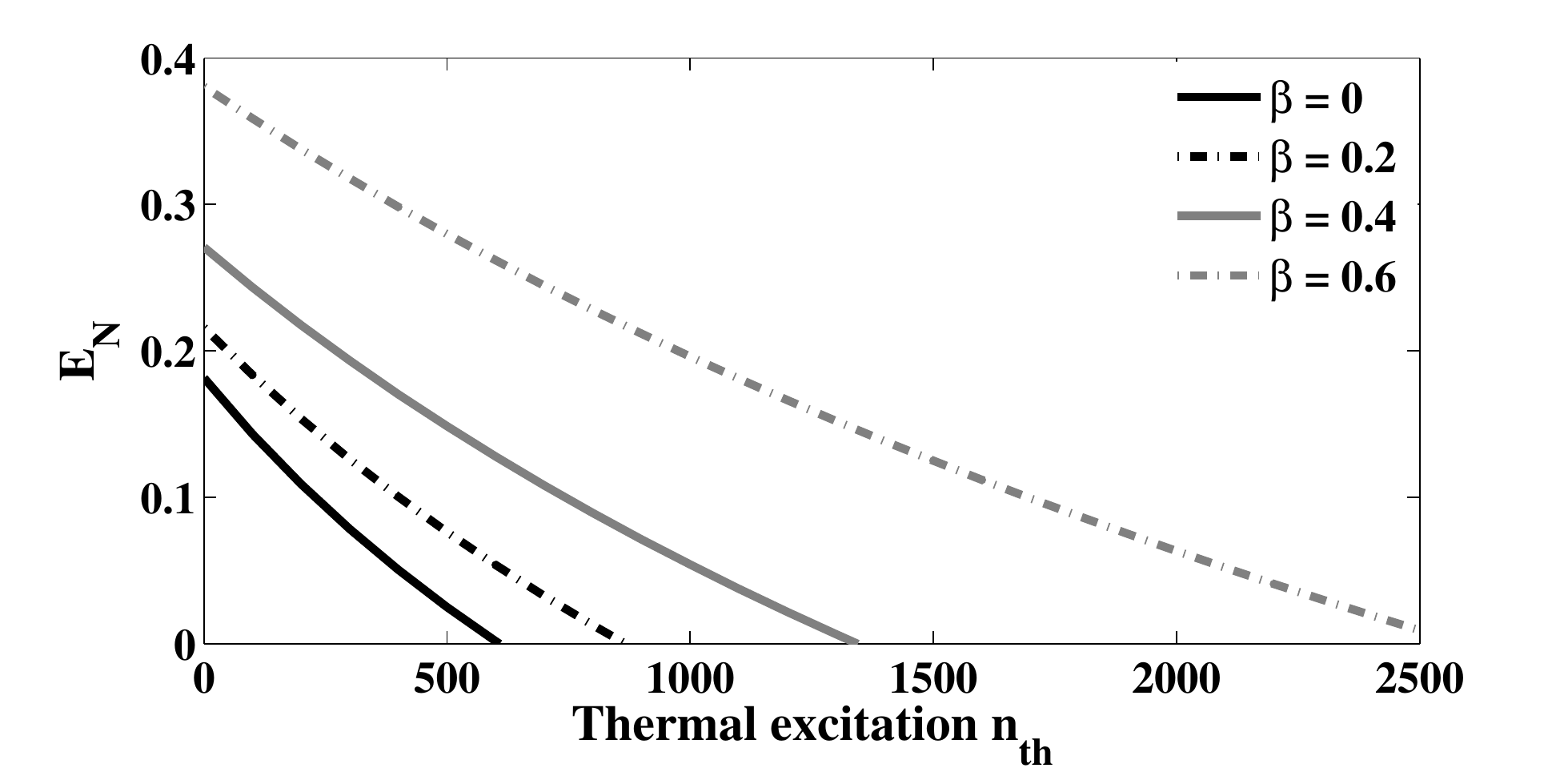}}
\caption{Plot of the logarithmic negativity versus the mean bath occupation for
different values of the nonlinear parameter with an input power 
$P_0=10\opn{mW}$. This shows the robustness of entanglement against thermal 
decoherence depending on the nonlinear term. For $\beta=0.6$ this robustness 
persists to $n_{th}=2500$ (gray dash-dotted line).}
\label{fig:Fig3} 
\end{center}
\end{minipage}    
\end{figure*}

Finally, Fig. \ref{fig:Fig3} shows the robustness of the entanglement against 
the environmental temperature in the presence of such nonlinearity. As expected, 
in the linear regime ($\beta=0$), there is a decrease of the entanglement with 
respect to the bath temperature. However, the entanglement persists until 
$n_{th}=600$ or $T=98\opn{K}$, which is several orders of magnitude larger than 
the temperature needed to cool a mechanical resonator to its quantum ground 
state. This shows the performance of the system studied here to resist thermal 
decoherence. Figure \ref{fig:Fig3} mostly reveals that by increasing the 
geometrical nonlinearity, one definitely improve the robustness of the 
entanglement against thermal decoherence in an optomechanical system. Indeed, 
the gray dash-dotted line shows that entanglement persists until $n_{th}=2500$, 
which corresponds to room temperature. It should be noted that similar results 
have recently been found in \cite{20} and \cite{26} with the inverse bandwidth 
and quantum optomechanical interface, respectively. Particularly in \cite{26}, 
where the hybrid system is studied, the entanglement remains strong even for 
$n_{th}=10^4$, which is greater than that found here ($n_{th}=2500$). This may 
be understood by the fact that hybrid systems combine two optomechanical 
couplings, optical and microwave couplings, to drive a nanoresonator. 

Since the physical parameters used here have been recently studied 
experimentally and they satisfy the system's stability condition for the 
feasible input power of $P_0=10\opn{mW}$, our results can then be implemented. 
This techniques of entanglement that used the geometrical nonlinearity is 
further promising to observe the quantum effects even for high environmental 
temperatures. This could lead to the improvement of the robustness of quantum 
teleportation protocols and other quantum applications. 

\section{Conclusion}
\label{sec:Concl}

In this work, we have shown that geometrical nonlinearity can be used to enhance
the entanglement between the optical intracavity field and the mechanical mode 
in optomechanical cavities. This nonlinear parameter enables the robustness of 
the entanglement against thermal decoherence, even at room temperature. With
experimental parameters close to those of a recently performed experiment
\cite{1}, entanglement persists until $n_{th}=2500$ for an input power of
$10\opn{mW}$. This scheme offers promising perspectives to produce robust light
entangled states for improving quantum information applications such as 
continuous-variables dense coding, quantum teleportation, quantum cryptography
and quantum computational tasks.


\end{document}